\pdfoutput=1

\documentclass[aps,prb,reprint,showpacs,superscriptaddress,byrevtex]{revtex4-1}

\usepackage{graphicx}
\usepackage{graphics}
\usepackage{amsmath}
\usepackage{amssymb}
\usepackage{amsfonts}
\usepackage{dcolumn}
\usepackage{dsfont}
\usepackage{latexsym}
\usepackage{rotating}
\usepackage{color}
\usepackage{latexsym}
\usepackage{bbm}
\usepackage{subfigure}
\usepackage{float}
\usepackage{epsfig}
\usepackage{epsf}
\usepackage{psfrag}
\usepackage{bm}
\usepackage{amsthm}
\usepackage{eucal}
\usepackage{mathrsfs}
\usepackage{url}
\usepackage{braket}
\usepackage{natbib}

\usepackage{color} 

\usepackage{hyperref}
\hypersetup{
colorlinks=true,final=true,
        linkcolor=blue,
        citecolor=blue,
        filecolor=blue,
        urlcolor=blue,
}

\begin{document}
	\title{$CrI_{3}$-$WTe_{2}$: A Novel Two-Dimensional Heterostructure  as Multisensor for $BrF_3$ and $COCl_2$ Toxic Gases.}
	 
	 \author{Amreen Bano}
	 \affiliation{Department of Physics, Barkatullah University, Bhopal-462026, India} 
	\author{Jyoti Krishna}
	\affiliation{ Department of Physics, Indian Institute of Technology Roorkee, Roorkee-247667, Uttarakhand, India}
	\author{T. Maitra}
	\affiliation{ Department of Physics, Indian Institute of Technology Roorkee, Roorkee-247667, Uttarakhand, India}
	\author{N. K. Gaur}
	  \affiliation{Department of Physics, Barkatullah University, Bhopal-462026, India}
	  
	\begin{abstract}
	 
	A new multisensor (i.e. resistive and magnetic) $CrI_{3}$-$WTe_{2}$ heterostructure (HS) to detect the toxic gases $BrF_3$ and $COCl_2$ (Phosgene) has been theoretically studied in our present investigation. The HS has demonstrated sensitivity towards both the gases by varying its electronic and magnetic properties when gas molecule interacts with the HS. Fast recovery time ($< 0.14 fs$) under UV radiation has been observed. We have considered two configurations of $BrF_3$ adsorbed HS; 1) when F ion interacts with HS (C1) and 2) when Br ion interacts with HS (C2). In C1 case the adsorption energy $E_{ad}$ is observed to be -0.66 eV while in C2 it is -0.95 eV. On the other hand in case of $COCl_2$ $E_{ad}$ is found to be -0.42 eV. Magnetic moments of atoms are also found to vary upon gas adsorption indicates the suitability of the HS as a magnetic gas sensor. Our observations suggests the suitability of $CrI_{3}$-$WTe_{2}$ HS to respond detection of the toxic gases like $BrF_3$ and $COCl_2$. 
	 
	 \end{abstract}

	\maketitle

	\section{Introduction}
Gas detection for environmental monitoring has innumerable applications in field such as industries and agriculture, medical diagnosis, military etc.\cite{yang,askuch} that utilizes the adsorption of gas molecules over materials. For the great technological perspective, it necessitates the material to have superior physical and chemical stability as well as the accessibility for chip-scale miniaturization of sensing elements for the low cost. Owing to the advent in 2D materials and increasing mass-market applications, the research in the gas sensors field have elevated rapidly due to continuing need for the highly sensitive, selective and faster response and recovery dynamics towards gas adsorption. The first 2D atomic crystal graphene has for a long time enticed because of its extraordinary mechanical and electronic properties. The desired requirement of high surface area, carrier mobility, chemical and thermal stability with low electronic temperature noise, power consumption and higher response time promises graphene to be used in the next generation devices employed in gas sensing and bio-sensing\cite{a,b,c,d,e,f,g,h,i,j,k}. Since each atom in graphene is a surface atom, it results in the ultrasensitive sensor response. It has been seen that the epitaxially grown graphene based sensors are ultrasensitive towards $NO_{2}$ gas\cite{pearce}. However, pristine graphene limits its potential upon physical adsorption of common gas molecules\cite{l,n,o,p} because of no dangling bonds. Thus for the chemisorptive enhancements, the surface is functionalized through polymers or metallic coating\cite{lang, pumera}. Other forms of the graphene like graphene oxide (GO) or reduced GO do serve as a dynamic material for high performance molecular sensors\cite{schedin}. Inspired by the performance of first 2D material, the gas sensing communities captured several hundreds of different 2D materials including elemental allotropes such as silicene, germanene borophene etc., and compound like transition metal dichalcogenides (TMDs)\cite{q}. These have been tremendously successful in detecting even the traces of gas molecules like $NO_{2}$, $SO_{2}$, $NH_{3}$ etc.\cite{wei,byu,lian,he}. The forte of these materials are their ability to engineer artificial heterostructures (HS). Because of the van der Waal interactions between the HS the lattice mismatching is not there that ultimately minimize the interfacial damages and chemical modification\cite{zhong}.\\
Recently the integration between magnetic layer and semiconductors initiate a new generation of advanced functional materials. By manipulating the exchange interactions the electronic structure in 2D materials can be altered\cite{hau,qiao,r}. Generally, the gas sensing mechanism is based on the principle of change in electronic properties with gas adsorption. Variation in magnetic properties upon gas adsorption has never been realized. We have investigated here the gas sensing ability of a magnetic HS with the magnetic $CrI_{3}$ integrated over $WTe_{2}$ monolayer upon interaction with noxious gases $BrF_{3}$ and phosgene ($COCl_{2}$). The $BrF_{3}$ a hazardous gas used mainly in processing of nuclear fuel. It is corrosive to metals and tissues and irritates the respiratory upon inhalation. On the otherhand, phosgene is highly toxic gas used in industries for production of pesticides and its immediate reaction starts even below 2-3 ppm. So far no investigation has been done on $BrF_{3}$ gas adsorbtion on sensor layer.\\
Thus this paper focusses on the study of how the gas molecules ($BrF_{3}$ and $COCl_2$) interfere with the electrical and magnetic properties upon interaction. We have also investigated the nature of adsorption and selectivity towards each gaseous molecules. Practically, a sensor's recovery time ($R_{T}$) is crucial for technological applications, thus $R_{T}$ for the highly selective gas molecule is calculated for this system.

\section{Computational Details}
At ambient temperature and pressure conditions the crystal structure of $CrI_{3}$-$WTe_{2}$ HS is shown in Fig 1(a). The results presented here are obtained using fisrt-principles approach which based on density functional theory \cite{dft} as implemented in Quantum Espresso package \cite{qe}. Ideally, $CrI_3$ exists in two crystal structures: 1) $AlCl_3$ type monoclinic array and 2) $BiI_3$ type rhombohedral   order \cite{michael}. Here we report our findings for monoclinic assembly of $CrI_3$ deposited over hexagonal structure of $WTe_2$ \cite{zhong}. In order to explore the electronic structure of pure and $BrF_3$/Phosgene gas adsorbed $CrI_{3}$-$WTe_{2}$ HS we have employed plane-wave ultrasoft pseudopotential method to trace the valance electron interactions. To serve the exchange-correlation potential, generalized gradient approximation (GGA) of Perdew-Burke-Erzernhof (PBE)\cite{pbe} has been implemented. A supercell of $2\times 2\times 1$ has been used to construct $CrI_{3}$-$WTe_{2}$ HS. The cut-off kinetic energy of 760 eV has been applied with $7\times 7\times 1$ K-mesh for Brillouin zones sampling. We have used these values after complete optimization process. To avoid any interaction among atomic orbitals we have provided a large vacuum of 17$\AA$ along z direction. The $CrI_{3}$-$WTe_{2}$ HS has been allowed to fully relax under the convergence of total energy and total forces which are found to be better than 1.0 meV. For gas sensing calculations, we have kept the structural geometry of $CrI_{3}$-$WTe_{2}$ HS fixed and periodically moved the gas molecules $BrF_3$ and Phosgene $COCl_2$ (one at a time) along z direction in order to acquire the equilibrium distance $d_{eq}$ between the HS and gas molecule. In case of $BrF_3$ gas molecule we have studied its interaction with the HS along two different orientations 1) F atom is interacting with HS surface and 2) Br atom is interacting with HS surface. The value of $d_{eq}$ obtained in $BrF_3$ in case 1 is $\sim$ 2.25$\AA$  whereas in case 2 it is observed to be $\sim$ 2.04$\AA$. Moreover in case of Phosgene gas molecule $d_{eq}$ is found within the range of $\sim$2.32$\AA$ to $\sim$2.34$\AA$. The adsorption energy of gas molecules $BrF_3$ (in both cases) and Phosgene adsorbed over $CrI_{3}$-$WTe_{2}$ HS was defined as:
\begin{equation}
E_{ad} = E_{molecule/CrI_3-WTe_2 HS} - E _{CrI_3-WTe_2 HS} - E_{molecule}
\end{equation}
where $E_{CrI_3-WTe_2 HS}$ and $E_{molecule}$ indicates the total ground state energy of HS and gas molecule before adsorbtion take place respectively and $E_{molecule/CrI_3-WTe_2 HS}$ shows the total ground state energy of gas molecule adsorbed HS.  
\begin{table}[]
	\centering
	\caption {Comparison of experimental and theoretically calculated bond length (\AA) of parent compounds $WTe_{2}$ and $CrI_{3}$ in HS. The bond lengths from theoretical structure is lesser as compared to experiments which shows an overall compression in the HS. The net compression is due to the formation of interfacial bonds among the Te and I ions upon optimization.}
	\
	\renewcommand{\arraystretch}{2}
	\label{tab1}
	\begin{tabular}{||c|c|c||}
		\hline\hline
		& \textbf{Exp (\AA)}& \textbf{Calculations (\AA)} \\
		\hline
		W-W\cite{srep} & 3.6 & 3.2 \\
		\hline
		W-Te\cite{srep} & 2.769 & 1.448 \\
		\hline
		Cr-Cr\cite{michael} & 3.96 & 2.16 \\
		\hline
		I-I\cite{michael} (axis) & 3.86 & 2.04 \\
		\hline
		Cr-I\cite{michael} & 2.72 & 1.43 \\
		\hline\hline
	\end{tabular}
\end{table}

\section{Results and Discussion}
In the present investigation we have studied the HS which is comprised of monolayer of $CrI_{3}$ deposited over the honeycomb $WTe_{2}$ monolayer. Because of the presence of magnetic $Cr^{3+}$ ion in $CrI_{3}$ layer of HS, we have first carried out calculation for two different magnetic configurations namely, ferromagnetic (FM) and antiferromagnetic (AFM) spin states at Cr site. Since the AFM state gives higher energy as compared to FM one, thus FM configuration is the stable magnetic state which is in accordance with the previous reports\cite{michael}. Hence, the further investigations have been done for FM configuration only.

\subsection{Structural Analysis}
The pristine HS shown in Fig 1(a) is composed of $WTe_{2}$ and $CrI_{3}$ monolayers. From Fig 1(a) we can see that in $WTe_{2}$, the W ions forms a zig-zag pattern along a-axis resulting in slightly distorted hexagonal symmetry. The Te ions constitute an octahedral environment accompanied with strong intra-layer covalent bonding w.r.t W ions. Whereas, in the $CrI_{3}$ layer of HS, $Cr^{3+}$ ions form a honeycomb lattice. The I- ions create an edge sharing octahedrally coordinated network w.r.t. $Cr^{3+}$ ions such that the three I- ions are coordinated at the top and bottom layer of Cr ions. The two parent compounds ($WTe_{2}$ and $CrI_{3}$ monolayer) are vertically stacked together along c-axis to form a $CrI_{3}$-$WTe_{2}$ HS with interfacial bonds linking I and Te ions. The $\langle Te-I \rangle $ average bond length at the interface is 2.61 \AA. An overall compression along c-axis has also been observed among the parent compounds of the HS which may affect its electronic structure. Table 1. displays the comparison of experimental and calculated bond length in HS. 
	\begin{figure}[h]
	\centering
	\includegraphics[width=0.5\textwidth]{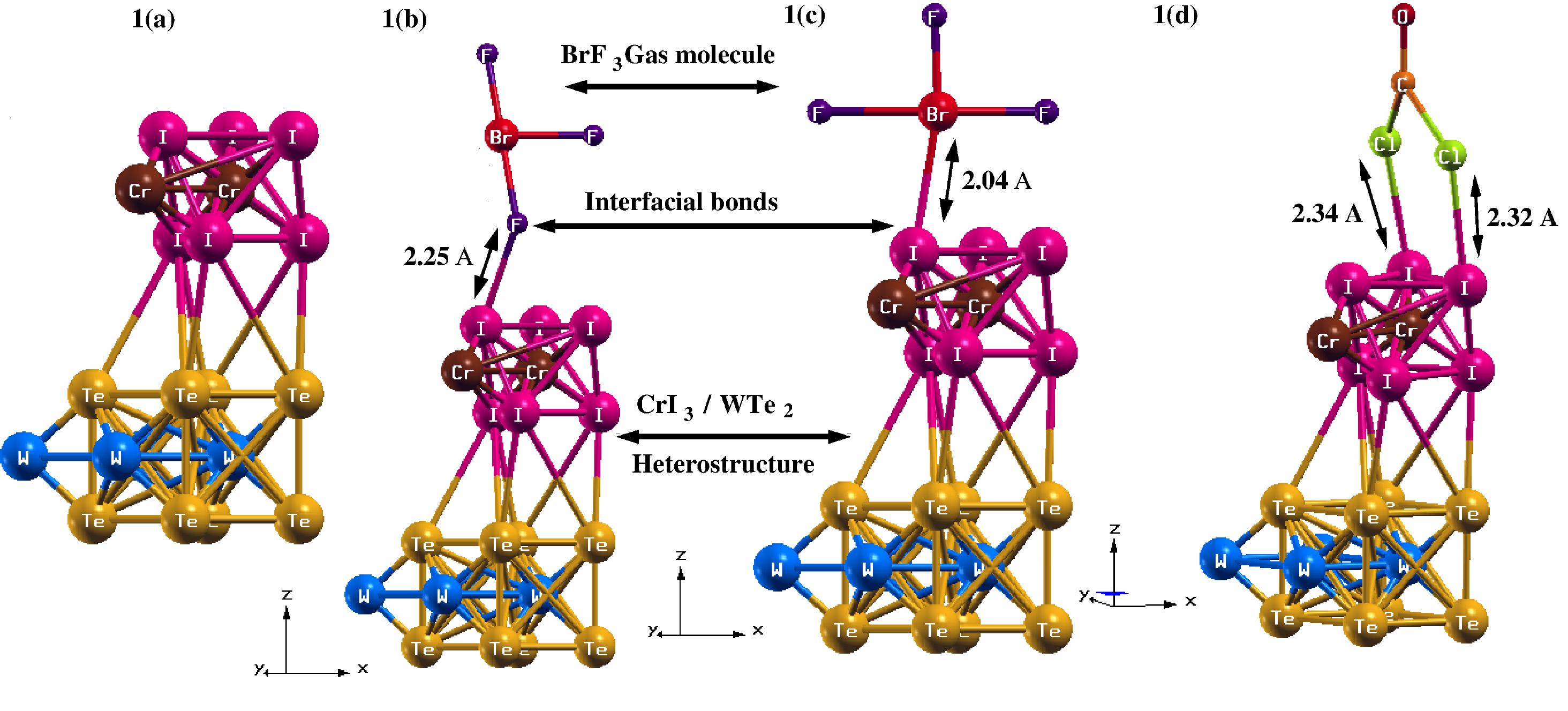}
	\caption{\label{fig: 1} Crystal structures of \textbf{a)} Pristine $CrI_3-WTe_2$ HS showing interfacial bonds among the parent compounds i.e. $CrI_3$ and $WTe_2$. These bonds results in net compression of the optimized HS. \textbf{b)} $BrF_3$ adsorbed $CrI_3-WTe_2$ HS with C1 configuration when F ion is directly interacting with the surface of the HS with a seperation of 2.25 \AA. \textbf{c)} $BrF_3$ adsorbed $CrI_3-WTe_2$ HS with C2 configuration when Br ion is directly interacting with the HS. The value of $d_{eq}$ in this case has been observed to be 2.04 \AA. \textbf{d)} $COCl_2$ adsorbed $CrI_3-WTe_2$ HS with $d_{eq}$ in range 2.32\AA to 2.34\AA. Only one configuration is considered for $COCl_2$ gas molecule due to larger reduced mass of Cl as compared to O ion.}
\end{figure}  

From the results obtained in Table 1 we observed a net compression in the HS (i.e. $<$ 45\%) 
except for $\langle W-W \rangle$ ($\sim$ 11\%) due to its relatively heavy atomic mass which obstructs any significant variation in its bond length as compared to other atoms. Hence the compression is emerging due to the interfacial bonds formed among the parent compounds of the HS (i.e. $WTe_{2}$ and $CrI_{3}$). These bonds are occuring from the charge transfer from Te ions to I ions (\textit{charges flows from low electronegativity (Te= 2.1 Pauling scale) to high electronegativity (I= 2.6 Pauling scale)}). This process of bond formation at the interface of HS in turn results in compression of bond length among the atoms upon optimization. The electronic properties of the HS may get influenced due to this compression which has been discussed in detail in the following section. The interaction of $BrF_{3}$ on HS can occur through two possible orientations: by forming an interfacial bond between (1) F and HS as shown in Fig 1(b) (C1 configuration), (2) Br and HS as shown in Fig 1(c) (C2 configuration). The $\langle F-Br-F \rangle $ bond angle is $86^{\circ}$  with the $\langle Br-F \rangle$ bond length along axial and equitorial plane as 1.72 {\AA} and 1.81 {\AA} respectively. The equilibrium distance ($d_{eq}$) in C1 and C2 case is 2.25 {\AA} and 2.04 {\AA} respectively. For the $COCl_{2}$ gas the bond angle and bond length is $124^{\circ}$ $\langle Cl-C-O \rangle$ and 1.76 \AA $\langle Cl-C \rangle$, 1.19 \AA $\langle C-O \rangle $ respectively. Unlike $BrF_{3}$, only single orientation of $COCl_{2}$ has been considered (Cl linked with HS) as presented in Fig 1(d) due to larger reduced mass of Cl w.r.t O (about $\sim$ 10\%) ions. The $d_{eq}$ varies from 2.32-2.34 {\AA} for this case.

\begin{figure} [h]
	\includegraphics[width=0.5\textwidth]{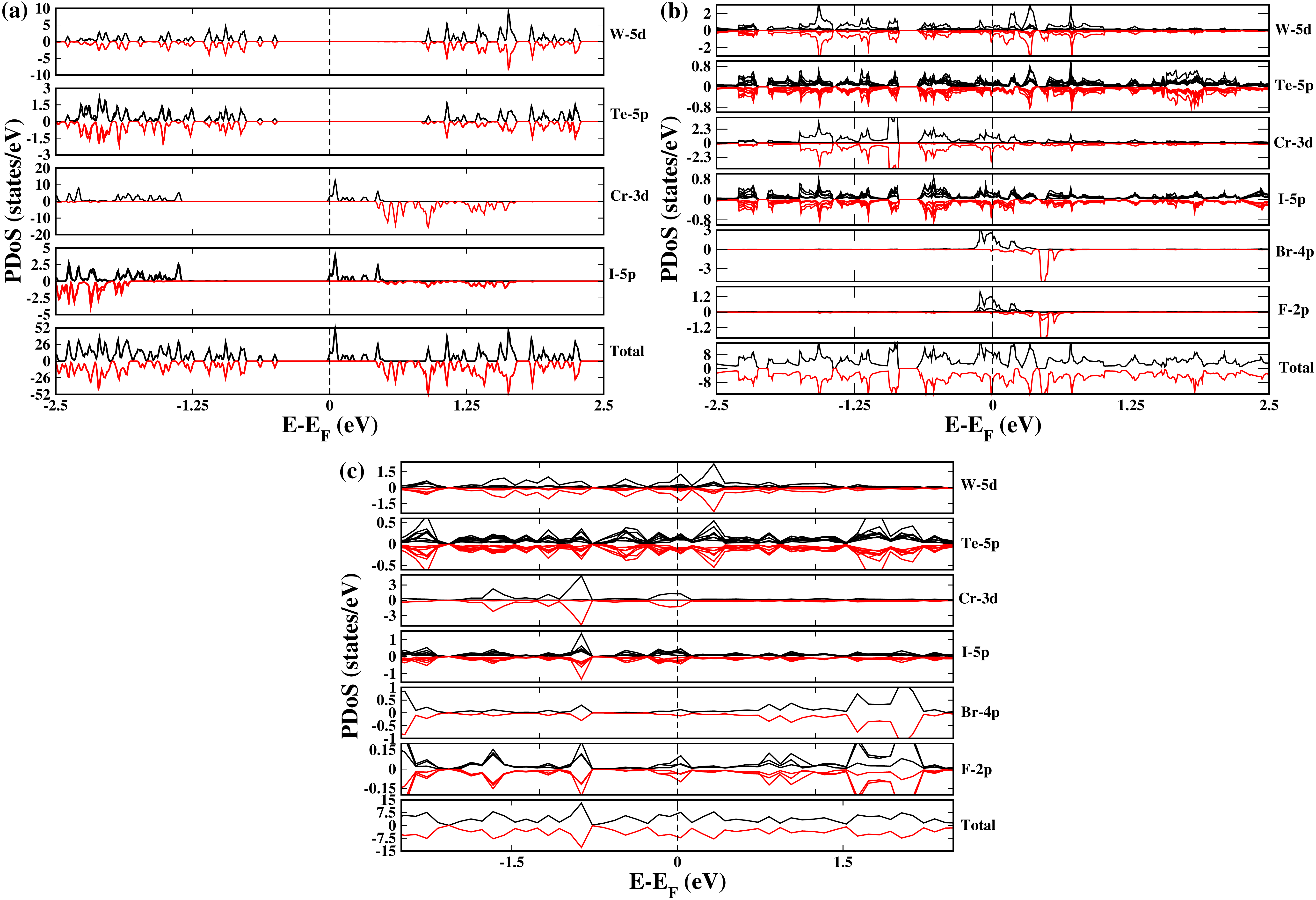}
	\caption{Results of Partial density of states (PDoS) of \textbf{a)} Pristine $CrI_3-WTe_2$ HS. The total DoS of HS showing half-metallic nature which is coming from the Cr$^{3+}$ ions having vacant 3d states. W-5d states and Te-5p states of $WTe_2$ monolayer is showing an insulating character with clear band gap of 1.31 eV. \textbf{b)} C1 of $BrF_3$ adsorbed HS. It is evident from the result that in C1 configuration the HS has become a metal unlike pristine HS which is a half-metal. On the other hand the spin-up states of the gas molecule $BrF_3$ is actively contributing in the metallicity of the HS while spin-down states are present around 0.58 eV in the conduction band region. \textbf{c)} C2 configuration of $BrF_3$ adsorbed HS. It can be seen that when Br ion interacts with HS there is no spin splitting exists among both spin channels unlike the C1 case.}
\end{figure}

	  \subsection{Electronic Structure}
	  In order to investigate the gas sensing effect of the HS, we have first studied the electronic density of states (DOS) prior to the gas adsorption. When no gas molecules were adsorbed, the total DOS of HS (Fig 2(a)) shows a spectral weight of 11.83 states/eV at Fermi level (FL). Small amount of metallicity is induced because of Cr-3d and I-5p orbitals of $CrI_{3}$ layer of HS. This induced metallicity is emerging from the compressed bond length of the atoms upon optimization as discussed above. Whereas the $WTe_{2}$ counterpart displays an insulating behaviour with a gap of 1.31 eV between majority and minority spin channels. Experimentally, $CrI_{3}$ layer is insulating in nature \cite{michael} but in $CrI_{3}$-$WTe_{2}$ HS, half-metallicity is observed. This might be due to the electron doping of $CrI_{3}$ layer induced by $WTe_{2}$ as reported previously \cite{wang1}. Near FL (E = -0.48 eV) only contributions from W-5d orbital and Te-5p orbital dominates whereas Cr-3d and I-5p orbital state lies 1.32 eV below FL. For pristine HS the total bandwidth for metallic state is observed to be 0.02 eV (Fig 2(a)) with the majority spin carriers separated from minority spin carriers by 0.88 eV. 
	 
	   \begin{figure} [h]
	   	\includegraphics[width=0.5\textwidth]{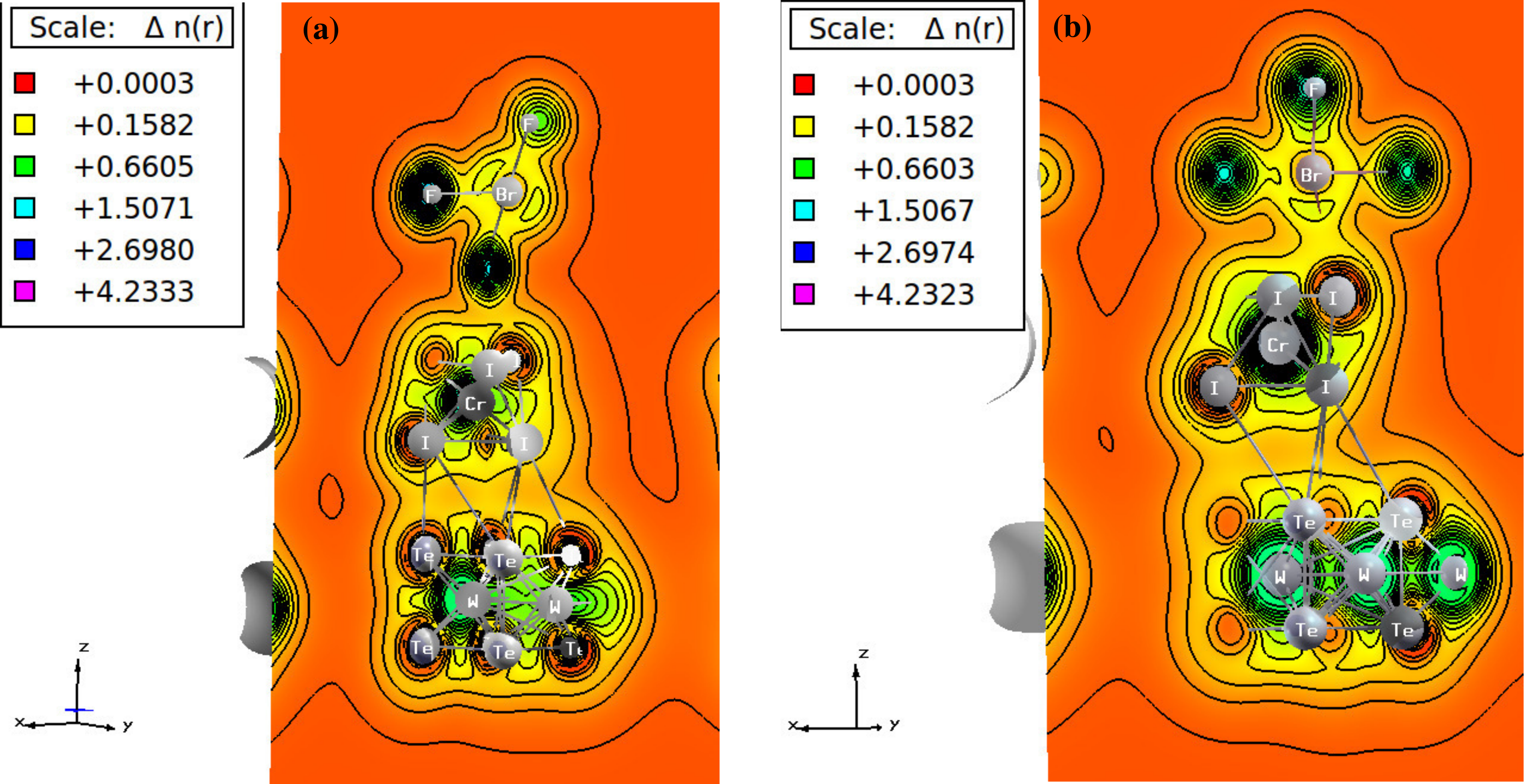}
	   	\caption{The charge density results of \textbf{a)} C1 configuration of $BrF_3$ adsorbed HS showing the charge transfer network among the HS and gas molecule. This also confirms the presence of chemisorptive nature of bond which has influeced the electronic structure of pristine HS. \textbf{b)} C2 configuration of $BrF_3$ adsorbed HS. The charges are appeared to flow from I ions to the gas molecule. Among all the atomic species presnt in the HS and the gas molecule, maximum electronegativity is possessed by F ions hence charges will get accumulate at F ions site.}
	   \end{figure}

	  \subsubsection*{1) $BrF_3$ Adsorbed HS:} When $BrF_3$ gas molecule is adsorbed on the HS the metallicity is enhanced in both the orientations which can be seen from Fig 2(b) and Fig 2(c). In C1 configuration (Fig 2(b)), when F directly forms an interfacial bonding with HS, the bandwidth increases to 0.67 eV. At FL, the dominant contribution is coming from Cr-3d states with the spectral weight for up and down spin density being 0.85 states/eV and 1.56 states/eV respectively. Feeble participation of W-5d (up 0.33 states/eV, down 0.77 states/eV) and Te-5p (up 0.14 states/eV, down 0.21 states/eV) and I-5p states(up 0.23 states/eV, down 0.37 states/eV) are also observed at FL. The adsorbed $BrF_3$ gas molecule in C1 have enhanced spin up DOS at FL. In principle, the charges should flow from Br (low electronegativity) to F (high electronegativity) ions but due to the large $\langle Br-F \rangle$ bond length ($<$ 1.7 \AA) the charge hopping takes place at slower rate resulting in higher spectral weight of Br (2.08 states/eV) ion as compared to F (0.97 states/eV) ions at FL. On the other side for C2 configuration (Fig 2(c)), when Br interacts directly with HS, the bandwidth further intensifies to 0.76 eV at FL. Likewise in C1, here also the Cr-3d states are pronounced at FL with spectral weight of 1.25 states/eV for all spin channels. Relatively weak involvement of W-5d (0.95 states/eV), Te-5p (0.2 states/eV) and I-5p (0.42 states/eV) states are present at FL. Now since the electronegativity of I, Br and F are 2.66, 2.96 and 3.98 Pauling scale respectively, therefore a continuous flow of the charge will take place from I to F via Br ion. Hence, net charge density at Br site decreases as compared to previous case. In C2, the net charge transport takes place through I-Br-F chain which facilitates the smooth transfer without any accumulation while the charges stocked at Br site due to uneven path (I-F-Br) w.r.t. electronegativity in C1. To investigate the nature of bonding between $BrF_{3}$ and HS, we have studied the charge density for both the orientations (Fig 3(a) and Fig 3(b)). For this purpose the calculations were performed in (1 1 0) plane. As discussed earlier that the charge transport channel is decided by the electronegativity difference. In C1 (Fig 3(a)), electronegativity of $WTe_{2}$, $CrI_{3}$ and $BrF_{3}$ is 1.5071, 2.6980 and 4.2333 respectively where the maximum charge is accumulated near F (4.2333) ion of $BrF_{3}$. Hence, a net flow of charge will take place from $WTe_{2}$ to $BrF_{3}$ layer via $CrI_{3}$. Similar charge flow network is followed in C2 as shown in Fig 3(b) where electronegativity of $WTe_{2}$, $CrI_{3}$ and $BrF_{3}$ is observed to be 1.5067, 2.6974 and 4.2323. 
	  
	   \begin{figure} [h]
	  	\includegraphics[width=0.5\textwidth]{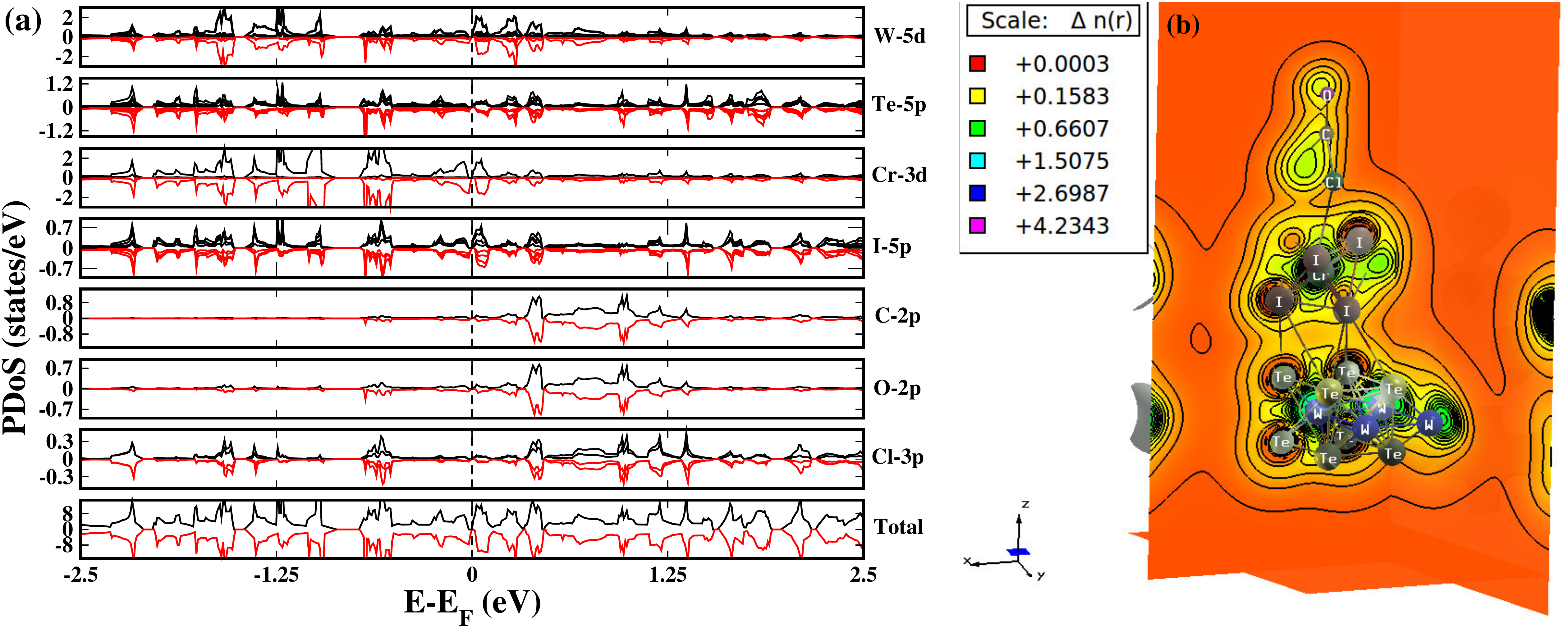}
	  	\caption{\textbf{a)} PDoS of $COCl_2$ adsorbed HS showing a 'gapless' type semiconducting behaviour. The term gapless is justified by the fact that the electronic states of HS species are not overlapped at the Fermi level. The energy states of gas molecule $COCl_2$ shows no significant contribution near Fermi level. \textbf{b)} The charge density result of Phosgene gas molecule adsorbed over HS indicates a uinidirectional charge flow from HS to the gas molecule which causes the charge accumulation at $COCl_2$. O ion is most electronegative in this case hence charges will be gathered at O ion site. }
	  \end{figure}
	  
	  \subsubsection*{2) Phosgene ($COCl_2$) Adsorbed HS:} Another poisonous gas $COCl_{2}$ when adsorbed on the HS surface exhibits a gapless type semiconducting behaviour which can be seen in Fig 4(a) showing the PDOS of the above stated system. With adsorption the states are present near the FL but does not crosses for W and Te ions. Similar situation is observed for Cr-3d and I-5d states. This is in contrast with $BrF_{3}$ adsorption where the metallicity was induced at each HS layer. Within -0.26 eV-0.31 eV energy window, the HS is occupied by the states of both spin channels near FL whereas, negligible contributions are coming from the gaseous states in that energy range. In the $COCl_{2}$ gas molecule, the electronegativity of C, Cl and O are 2.55, 3.16 and 3.44 Pauling scale respectively. Thus, the charge flow from C will take place to O and Cl ions resulting in slight occupation of states below FL in O and comparitively higher states at Cl site. The DOS of Cl ions (Fig 4(a)) has the 3p states peaked within 0.5 eV - 0.7 eV below FL due to higher charge density coming from C and HS layers. From the charge density calculations we observed a charge flow direction from HS to the gas molecule as given in Fig 4(b). The electronegativity of $WTe_{2}$, $CrI_{3}$ and $COCl_{2}$ is 1.5075, 2.6987, 4.2343 respectively. It results in a unidirectional charge flow from bottom of HS to the gas molecule which causes the accumulation of carriers at $COCl_{2}$. The above observations suggest that $CrI_{3}$-$WTe_{2}$ HS serve as a potential gas sensor for $BrF_{3}$ and $COCl_{2}$ gas molecule.
	   
	  \subsection{Magnetic Properties} In HS the magnetic contribution is coming due $CrI_{3}$ layer which has FM ordering with total spin magnetic moment of 5.35 $\mu_{B}$. And the magnetic moment per Cr and I ions are 2.86 $\mu_{B}$ and 0.041 $\mu_{B}$  respectively. This is in good agreement with the saturation moment(3.1 $\mu_{B}$/Cr) measured experimentally\cite{michael}. The low magnetic moment of I ions is due to the transfer of unpaired 4s electron from Cr to I- resulting in stable 5p states. This delocalization of charges causes reduced moment at I- site. On the other hand, the $WTe_{2}$ layer in HS remains non-magnetic. With the exposure of $BrF_{3}$ gas molecule in C1 configuration, the delocalization effects dominate in Cr ions resulting in decreased magnetic moment. The Br and F ions on the other hand acquire charges from HS have higher mag moment ( 0.088 $\mu_{B}$ for Br and 0.005 $\mu_{B}$, 0.144 $\mu_{B}$, 0.0268 $\mu_{B}$ per F ion)  as compared to Cr$^{3+}$ (0.009 $\mu_{B}$). As discussed above that due to difference in $ \langle Br-F \rangle $ bond lengths uneven charge flow takes places to form dissimilar magnetic moment per F ion. In C2 configuration when Br directly bonded with HS layer total magnetic moment is almost negligible. This is in accordance with the DOS of C2 (Fig 2(c)) where the net reduction in spectral weight was observed. Due to the continuous charge transfer path (I-Br-F) the delocalization of electrons causes the moments to drop. The same scenario has been observed when Phosgene is exposed to HS. The overall reduction in magnetic moment is observed here as well. The variation in magnetic moments of the HS upon interaction of gas molecules suggests that $CrI_{3}$-$WTe_{2}$ HS can be used as a magnetic gas sensor as well as resistive gas sensor. There are many studies over the latter type but only few experimental studies are performed on the former type of gas sensor which detects the perturbation in the magnetic properties when gas molecules interact with the sensor material\cite{alex}. A few magnetic gas sensors so far studied are nanoparticles of $CuFe_2O_4$ which was used for the detection of volatile organic compounds (VOCs)\cite{mata}, Co/ZnO nanorods to detect $H_2$ and CO molecules\cite{pon}, Co/ZnO hybrid nanostructures for the detection of $C_3H_6O$, CO and $H_2$ target gases\cite{cip} etc.
	  
	   \subsection{Adsorbtion and recovery time}
	  The adsorbtion energy describes the nature of stability among adsorbent (HS) and adsorbate (gas molecule). The process can take place in two modes (1) physisorption: which involves weak van der Waals forces between two reacting species. The electronic properties of adsorbent is barely perturbed during this mechanism; (2) chemisorption: here actual involvement of chemical bonds between species exists. This also require minimum activation energy to initiate the process. In C1 with $BrF_{3}$ adsorption on HS, the adsorption energy is -0.66 eV while in C2 it increases to -0.95 eV. The increasing adsorption rate by 30\% suggests the comparatively strong chemisorptive nature in C2. The existence of strong covalent bonding between HS and $BrF_{3}$ (C1 and C2) have been observed from charge density results shown in Fig 3(a) and Fig 3(b). Similar studies for $COCl_{2}$ adsorption shows the chemisorptive character but the $E_{ad}$ (-0.42 eV) is relatively smaller than that from $BrF_{3}$ interaction. It is evident that as $d_{eq}$ increases, $E_{ad}$ energy decreases. Thus, the higher $d_{eq}$ for $COCl_{2}$ case is marked by decrease in $E_{ad}$. Though the relative stability in $COCl_{2}$ case is lesser than $BrF_{3}$ but from previous literature $COCl_{2}$ adsorption on BN nano tube (BNNT), BN nano rod (BNNR) and borophene reported to have $E_{ad}$ -0.18 eV, -1.058 eV and -0.306 eV respectively \cite{behe,khusboo}. Hence, $COCl_{2}$ adsorpion on $CrI_{3}$-$WTe_{2}$ HS has shown better performance as compared to previous reports with an exception of BNNR.\\
	  The recovery time $R_{T}$ of a sensor is based on how fast the sensor retrieve its initial state. Based on the Arrhenius theory the sensor recovery time \cite{pitt} is related by:
	  \begin{equation}
	  R_{T} = \nu^{-1}e^{-E_{ad}/KT}
	  \end{equation}
	  where, $\nu$ is the operational frequency, $E_{ad}$ is adsorbate energy, K is Boltzmann constant and T is the sensor's operational temperature. For different attempt frequencies, the sensor's recovery rate is affected as tabulated in Table 2. Under UV illumination the HS is showing faster $R_{T}$ for all the cases. The recovery rate depends on the nature of adsorption. With relatively weak chemisorptive effect of $COCl_{2}$ gas on HS, fastest recovery time is achieved.
	  
	  \begin{table}[]
	  	\centering
	  	\caption{Adsorbtion energy ($E_{ad} (eV)$) and recovery time $ R_{T}$ (fs) for $BrF_{3}$ and $COCl_{3}$ gas adsorption over HS layer. In the I coloumn, the $E_{ad}$ of $BrF_{3}$ and $COCl_{2}$ gases showing chemisorptive nature. The highest $E_{ad}$ in $BrF_{3}$ represents the strongest covalent bonding between the gas molecule and HS. From the II, III and IV coloumns, the recovery is fastest when the sensor is illuminated with UV radiation. }
	  	\
	  	\renewcommand{\arraystretch}{1.5}
	  	\label{tab1}
	  	\begin{tabular}{||c|c|c|c|c||}
	  		\hline\hline
	  		& E$_{ad}  (eV)$ & $R_{T}$ (IR) (fs) & $R_{T}$ (Visible) (fs) & $R_{T}$ (UV) (fs) \\
	  		\hline\hline
	  		\textbf{$BrF_{3}$  (C1)} & -0.66 & 1302 & 13.02 & 0.13 \\
	  		\hline
	  		\textbf{$BrF_{3}$ (C2)} & -0.95 & 1460 & 14.6 & 0.14 \\
	  		\hline
	  		\textbf{$COCl_{2}$} & -0.42 & 1180 & 11.8 & 0.12 \\
	  		\hline\hline
	  	\end{tabular}
	  \end{table}
	  
	  \section{Conclusions}
	  We have theoretically investigated a new 2-dimensional $CrI_3$-$WTe_2$ HS in the present work in order to explore the possibility as a multisensor (i.e. resistive and magnetic). Our results shows that upon interaction with the gas molecules $BrF_{3}$ and $COCl_{2}$ with HS the electronic as well as magnetic properties of pristine HS get altered. We have also determined that how swiftly the HS can get recover after detaching the gasous species from it by means of recovery time. We found that under UV illumination ultrafast recovery time is presented by the HS i.e. $< 0.14 fs$. Hence we conclude that $CrI_3$-$WTe_2$ HS offers it self as a multisensor for the detection of highly toxic gases like $BrF_{3}$ and $COCl_{2}$.

	\end{document}